\documentclass[letterpaper,twocolumn,10pt]{article}
\usepackage{usenix-2020-09}

\usepackage{tikz}
\usepackage{amsmath}
\usepackage{hyperref}
\usepackage[noblocks]{authblk}

\usepackage{biblatex}
\begin{document}

\date{}

\title{\Large \bf FlexBSO: Flexible Block Storage Offload for Datacenters}

\author{Vojtech Aschenbrenner}
\author{John Shawger}
\author{Sadman Sakib}
\affil{Department of Computer Sciences, University of Wisconsin-Madison}

\maketitle

\section{Introduction}

Efficient virtualization of CPU and memory is standardized and mature. Capabilities such as Intel VT-x~\cite{uhlig2005intel} have been added by manufacturers for efficient hypervisor support.
In contrast, virtualization of a block device and its presentation to the virtual machines on the host can be done in multiple ways.
Indeed, hyperscalers develop in-house solutions to improve performance and cost-efficiency of their storage solutions for datacenters.
Unfortunately, these storage solutions are based on specialized hardware and software which are not publicly available.

The traditional solution is to expose virtual block device to the VM through a paravirtualized driver like \texttt{virtio}~\cite{russell2008virtio}.
\texttt{virtio} provides significantly better performance than real block device driver emulation because of host OS and guest OS cooperation.
The IO requests are then fulfilled by the host OS either with a local block device such as an SSD drive or with some form of disaggregated storage over the network like NVMe-oF or iSCSI.

There are three main problems to the traditional solution.
1) \textit{Cost.} IO operations consume host CPU cycles due to host OS involvement.
These CPU cycles are doing useless work from the application point of view.
2) \textit{Inflexibility.} Any change of the virtualized storage stack requires host OS and/or guest OS cooperation and cannot be done silently in production.
3) \textit{Performance.} IO operations are causing recurring \texttt{VM EXITs} to do the transition from non-root mode to root mode on the host CPU.
This results into excessive IO performance impact.

We propose FlexBSO, a hardware-assisted solution, which solves all the mentioned issues.
Our prototype is based on the publicly available Bluefield-2 SmartNIC with NVIDIA SNAP support, hence can be deployed without any obstacles.

\section{Design and Implementation}

\begin{figure}[htbp]
  \centering
  \includegraphics[width=0.45\textwidth]{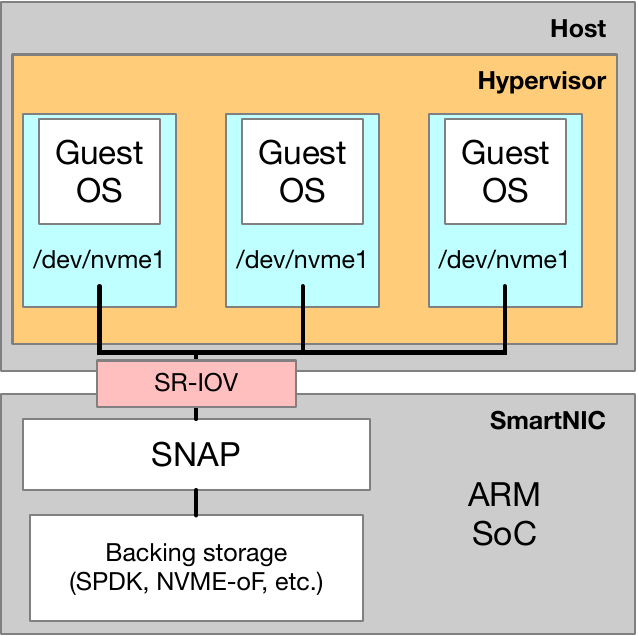}
  \caption{SR-IOV allows using FlexBSO from multiple VMs without host OS intervention.}
  \label{fig:sr-iov}
\end{figure}

FlexBSO uses the Bluefield-2 SmartNIC to completely replace the storage stack of the hypervisor and solves aforementioned issues of traditional paravirtualized solutions.
It uses NVIDIA SNAP with SR-IOV to directly expose an NVMe block device to every single guest on the host (Figure~\ref{fig:sr-iov}.
The guest works with the block device as if it was a local NVMe device, but all IO commands are fulfilled by the Bluefield-2 card.
The host OS is completely bypassed.

SNAP can be viewed as an SPDK storage stack enhanced with a subsystem presenting virtual NVMe devices on PCIe bus.
This enables huge flexibility of the solution, because the logic of the storage is just well-known SPDK~\cite{yang2017spdk}.
It is trivial to modify and/or create new SPDK block devices, to add more layers into the storage stack.
For example, if the customer wants to increase durability of his data, it is possible to seamlessly change the RAID mode from 0 to 1.
More features can be enabled in similar fashion, like encryption of compression.
The flexibility can be taken to the extreme, in fact, this solution enables changing the whole storage backend if necessary without any notice from the guest perspective.

Because the host OS is completely bypassed, this solution eliminates all \texttt{VM EXITs} caused by traditional solutions, which leads into significant performance boost and cost efficiency.

In the following sections, we describe used technologies in a higher detail and share our experience with their modifications.

\subsection{SNAP}
The proprietary SNAP (Software-defined Network Accelerated Processing) library from NVIDIA allows DPUs such as the Bluefield-2 to emulate an NVMe storage device on the host PCIe bus. The host can use standard NVMe drivers to interact with the device. SNAP can operate in two modes:
\begin{enumerate}
\item \textit{Fully-offloaded mode:} NVMe requests are sent directly to an NVMe-oF target by the SNAP accelerator. In this mode, the ARM cores on the DPU are used solely for the control plane and do not touch the data.
\item \textit{Partially-offloaded mode:} NVMe requests are sent to an SPDK storage stack running on the ARM cores of the DPU. This allows for greater flexibility in handling the data, potentially using encryption, compression, or other accelerators present on the DPU.
\end{enumerate}

\begin{figure}[h]
  \centering
  \includegraphics[width=0.45\textwidth]{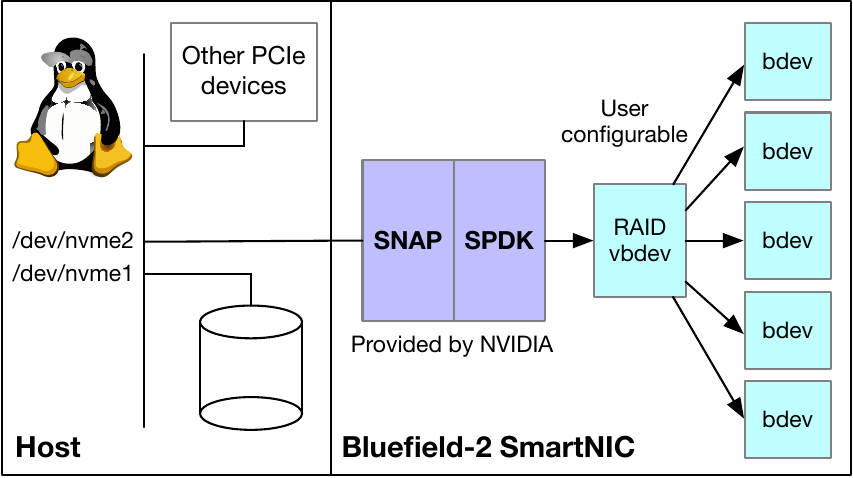}
  \caption{System diagram}
  \label{fig:system-diagram}
\end{figure}

The details of our system are shown in Figure~\ref{fig:system-diagram}. \texttt{/dev/nvme1} is a traditional NVMe disk on the host. \texttt{/dev/nvme2} is the NVMe device presented by SNAP. In this partially-offloaded configuration, the SNAP and SPDK components of the storage stack on the SmartNIC are provided by NVIDIA and should not be modified. The block device attached to SPDK is user-configurable, and can be replaced by any SPDK block device or NVMe-oF target. We found that it is possible to compile and link SPDK block devices against the SPDK library on the SmartNIC, allowing us flexibility in how the SmartNIC processes NVMe block requests.

\subsection{SPDK and Block Devices} \label{spdk}
The storage performance development kit (SPDK) is a user-level library and NVMe driver commonly used to build high performance storage applications. Traditional kernel-based I/O requires expensive context switches in response to hardware interrupts. SPDK operates storage drivers in polled-mode rather than interrupt-mode, vastly decreasing operation latency at the cost of using a core for continuous polling. We implemented two SPDK block devices to explore the flexibility of the system.

\subsubsection{RAID Block Device}
There are two types of block devices in SPDK -- virtual block devices (vbdevs) and terminal block devices (bdevs). Virtual block devices receive I/O submissions and do some computation before submitting I/O to other block devices. A classic example of a virtual block device is a RAID~\cite{raid} controller. SPDK is distributed with a RAID vbdev which is capable of RAID0 (striping), RAID1 (mirroring), and RAID5 (distributed parity), although it has several limitations. Notably, the RAID5 controller is only capable of full-stripe writes. It does not do read-modify-write required for single block writes in RAID5. We were primarily interested in exploring in the flexibility of the system, so this limitation did not concern us.

\begin{figure}[h]
  \centering
  \includegraphics[width=0.35\textwidth]{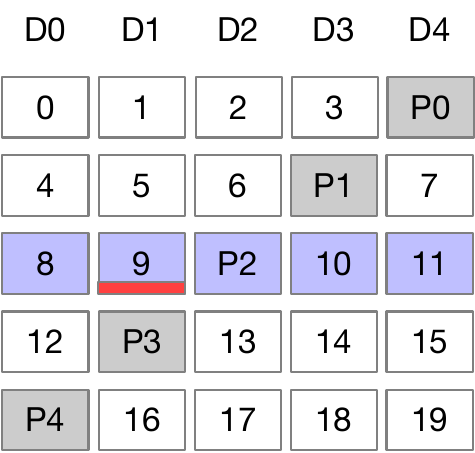}
  \caption{RAID5 ``safe read''}
  \label{fig:raid-blocks}
\end{figure}

We chose to implement a ``safe read'' in the RAID5 controller. To do so, we modified the provided RAID5 controller to perform a full-stripe read and recompute parity against the desired block to be read. If the recomputation does not match the original data, we notify the user that a parity check has failed. An example is shown in figure~\ref{fig:raid-blocks}. The red block refers to the data requested by the user. Our RAID5 will read in the entire strip (Blocks 8-11 including parity) and recompute the parity of block 9. Then, the red region of the block will be compared against the recomputed block to check for bit errors. To implement this, we reuse the \texttt{reconstruct\_read} function already present in the RAID5 driver. When the completion callback for a read I/O is executed, we do a \texttt{reconstruct\_read} on the read block's stripe, with the the read block as the reconstruct target. We also added functionality to ``poison'', or flip the first bit of, a write block with 0.1\% probability to help test this feature. 

\subsubsection{Compression Block Device}

To demonstrate a usecase of our flexible block storage device, we implemented a compression block device in the Bluefield-2 DPU. A compression block device is a virtual storage device on top of a physical storage device that can compress and decompress data transparently. This reduces the storage space required for the data.

We implemented the compression block  device in SPDK which contains a virtual block device (bdev) and a malloc block device as base. In the Bluefield-2 DPU, we used SPDK version 23.01. The virtual block device is implemented as a block device module, which is SPDK's equivalent to a device driver in an operating system. The module provides a set of function pointers that are called to service block device I/O requests. So, any other application can send I/O requests to this virtual block device using SPDK I/O submission functions such as \texttt{spdk\_bdev\_read()}. 

We used DOCA compress library to perform compression and decompression at the block device layer. DOCA compress library provides the API to compress and decompress data using hardware acceleration. It supports both host and DPU memory regions. Bluefield-2 device supports compression and decompression using the deflate algorithm.

The virtual bdev performs compression on write and decompression on read with doca compression library. The compression is done in the virtual block device before writing compressed data to base block device. Decompression is done in the read callback function of the virtual bdev which is invoked after base block device has finished reading. 

\section{Experimental Results}

We performed an experimental evaluation of crucial parts of the system.
First, we validate the feasibility of the solution in terms of throughput necessary for several VMs running on one node.
This is the most crucial result of the work, because having a bottleneck between the VM and the SmartNIC would be unacceptable.

Second, we explore the flexibility of our solution using SPDK block devices.
We share results from the process of customizing the storage stack on the SmartNIC.
Our experience was good enough to verify the flexibility of the solution.

\subsection{SNAP performance}

The crucial question is if FlexBSO will provide throughput sufficient enough to saturate needs of multiple VMs running on a single host.
To have a baseline, we compared FlexBSO to NVMe-oF using RDMA, which represents one of the high-performance solutions used traditionally.
Both solutions were configured with SPDK running on the SmartNIC.
FlexBSO uses SNAP to expose the NVMe block device to the VM and the baseline uses NVMe-of target provided by SPDK on SmartNIC and NVMe-of client on the VM.

\begin{figure}[htbp]
  \centering
  \includegraphics[width=0.45\textwidth]{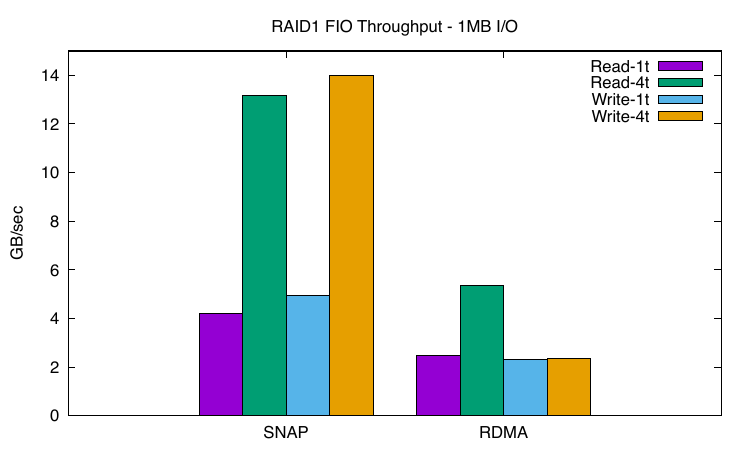}
  \caption{Throughput of exposed NVMe device via SNAP and NVMe over RDMA. The microbenchmark was performed by FIO with 1MB block size, 32 IO depth, direct IO, runtime of 60 seconds and 4 and 1 threads eventually. SPDK backend was configured as RAID1 device backed by Null block device. SNAP is capable of reaching up to 14GB/s in throughput, which makes SNAP a suitable solution, since many VMs can share this bandwidth without causing a bottleneck for common use cases.}
  \label{fig:tput}
\end{figure}

Figure~\ref{fig:tput} shows the throughput of both solutions. FlexBSO (SNAP) clearly dominates, and has more than $3\times$ higher throughput in multi-threaded scenario with the workload oriented for throughput. Another interesting metric is a latency of the solution. For a latency oriented read workload, i.e. single thread, io depth 1 and block size of 4kB, we measured read latency of $16\mu s$ for FlexBSO and $63.7\mu s$ for RDMA. This makes FlexBSO latency almost $4\times$ lower than RDMA.

\subsection{Custom SPDK bdevs}
Our first custom bdev, RAID5, was developed against a more recent version of SPDK than was present on the SmartNIC. When trying to compile our block device on the SmartNIC, we encountered linking errors. After further investigation, we found that the different versions of SPDK, although only about a year apart, had several interface incompatabilities. Furthermore, the RAID5 block device was significantly changed between versions. We tried to re-port our vbdev to SPDK on the SmartNIC, but we were not able to complete this work in the time available to us. Furthermore, SPDK on the SmartNIC was compiled without the flag to enable RAID5, so we were not able to test SPDK's default RAID5 device. We were reluctant to recompile SPDK on the SmartNIC, as SNAP must also be recompiled against custom versions of SPDK, as described in NVIDIA's documentation, and we did not have access to the SNAP source code.

\begin{figure}[h]
  \centering
  \includegraphics[width=0.35\textwidth]{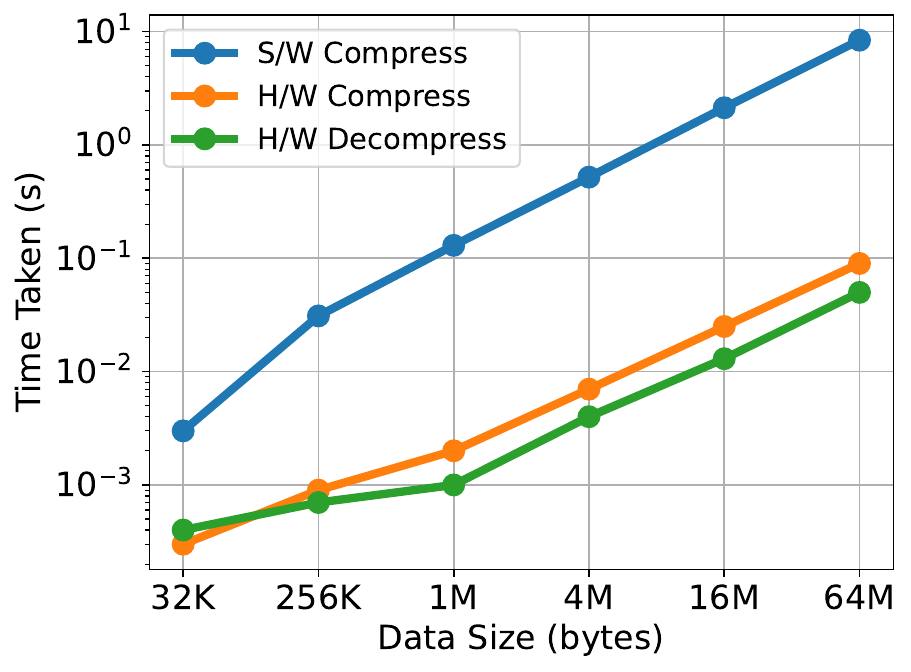}
  \caption{Time required to perform compression and decompression individually using zlib library and hardware acceleration}
  \label{fig:doca-compress}
\end{figure}

To understand the performance of our compression bdev, we first measured the time taken to perform (de-)compression in a single job in software using the zlib library, and in hardware using the DOCA compress library (Figure \ref{fig:doca-compress}). We used text data that maintained a compression ratio of about 4. The data size is the size of input data in compression and the size of output data in decompression. In both software and hardware methods, completion time increases linearly with the data size. Roughly, the hardware accelerated jobs took two order of magnitude less time then the software jobs on same data size. However, we saw error in hardware accelerated (de-)compression for data size greater than 128 MB. This is likely due to limitation of size of the DOCA buffer that can be allocated in the Bluefield-2 DPU.

\begin{figure}[h]
  \centering
  \includegraphics[width=0.35\textwidth]{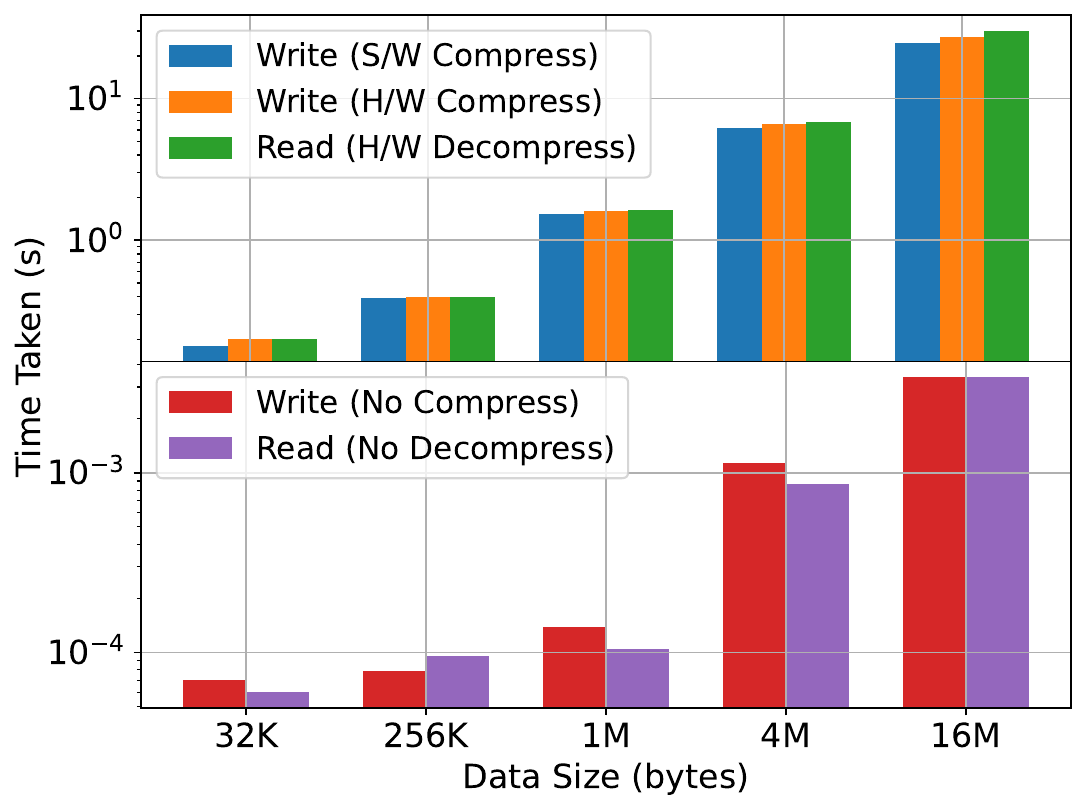}
  \caption{Time required to perform read and write operation through virtual block device with and without compression feature}
  \label{fig:bdev-compress}
\end{figure}

Next, we measured the time required to perform read and write operation through the virtual block device with and without compression feature (Figure \ref{fig:bdev-compress}). For data size greater than and equal to 128 KB, we used block size 128 KB which is the maximum allowed size in SPDK. The whole data was first written to the block device, and after write finished, it was read completely. Roughly, the I/Os with compression took four order of magnitudes more time than without compression. The (de-)compression in software and hardware took similar time and both likely experienced similar overhead. Following SPDK design, all IOs completed asynchronously on a non-blocking path. However, this required creating DOCA compress software context each time and using separate workqueue for each DOCA compress job. For larger data size, this likely adds significant overhead. We found memory errors in DOCA when performing IO with compression for data size more than 16 MB.

\section{Conclusion}
NVMe emulation using SNAP compares positively to existing remote-storage storage solutions such as NVMe-oF. We experienced nearly three times the throughput in a multi-threaded environment. We believe another benefit of SNAP is that it lets developers add flexibility to the block device abstraction, since it is software defined. Our experience suggests that doing so is feasible, however it requires a detailed knowledge of the SPDK environment and judicious use of computational resources on the SmartNIC. A promising next step for this work would be to investigate a multi-tenant environment on the host, using SR-IOV. In particular, we would be interested in the scalability of SNAP, and how many host VMs it is able to support efficiently.


\printbibliography

\end{document}